\documentclass[showpacs,amsmath,amssymb]{revtex4}

\usepackage{graphicx}
\usepackage{dcolumn}
\usepackage{bm}

\begin{document}


\title{Hysteresis behavior of the random-field Ising model with 2-spin-flip dynamics: Exact results on a Bethe lattice }

\author{Xavier Illa} 
\email{xit@ecm.ub.es}  
\affiliation{ Dept.   d'Estructura i  Constituents  de la  Mat\`eria,
Universitat  de Barcelona  \\ Diagonal  647, Facultat  de F\'{\i}sica,
08028 Barcelona, Catalonia, Spain}
\author{Martin-Luc Rosinberg }  
\author{Gilles Tarjus} 
\affiliation{Laboratoire de Physique Th\'eorique de la Mati\`ere Condens\'ee,
Universit\'e Pierre et Marie Curie, 4 Place Jussieu, 75252 Paris Cedex 05, France}
\date{\today}

\begin{abstract}
We present an exact treatment of the hysteresis behavior of the zero-temperature random-field
Ising model on a Bethe lattice when it is driven by an external field and evolved according to a 2-spin-flip dynamics. We focus on lattice connectivities $z=2$ (the one-dimensional chain) and $z=3$. For the latter case, we demonstrate the existence of an out-of-equilibrium phase transition, in contrast with the situation found with the standard 1-spin-flip dynamics. We discuss the influence of the degree of cooperativity of the (local) spin dynamics of the nonequilibrium response on the system.
\end{abstract}



\maketitle

\section{Introduction}
\label{Intro}

The nonequilibrium response of the random field Ising model (RFIM) at zero temperature to a smoothly changing external magnetic field has been introduced by Sethna et al.\cite{SDKKRS1993,PDS1995} to describe the hysteresis behavior of low-temperature magnets, in particular the property of ``return-point memory" and the power-law scaling characteristic of the ``Barkhausen noise" \cite{DZ2004}. In such a description, the competition between the quenched disorder and the ferromagnetic interactions induces a very large number of local energy minima in which the system gets trapped. In the zero-temperature dynamics, no escape by thermal activation is possible and evolution only results from changes in the driving field. Instead of wandering to find energetically more favorable states, the system then goes to the closest minimum made accessible by the field change. The state of the system is history-dependent and the evolution proceeds by ``avalanches",  i. e., jumps from one mini!
 mum to another. An extensive investigation of this zero-temperature hysteretic behavior of the RFIM has been carried out by several authors (for a recent review, see \cite{SDP2004}). It has been shown that the model displays an out-of-equilibrium critical point, with its associated scaling and universality properties. The critical point separates a strong-disorder regime in which  the magnetization curve and the corresponding hysteresis loop  is smooth on the macroscopic scale and a weak-disorder regime in which it shows a discontinuous jump at a nonzero value of the external field.

Real materials, however, are not at zero temperature. Application of the above picture requires a separation of time scales: the time for "local equilibration", namely the duration of the avalanche that follows a change in the applied field, must be much shorter than the experimental time scale, set by the rate of change of the applied field, which itself must be much less than the characteristic lifetime of the relevant metastable states in which the system is trapped. The extreme limit of the former condition is the "adiabatic" case in which the rate of change of the external field goes to zero. The extreme limit of the latter condition corresponds to the zero-temperature ($T=0$) dynamics in which no thermally activated escape takes place. The effect of a finite rate has been fairly extensively studied\cite{TWD2002,WD2003}; that of a finite temperature is a tougher issue.

To study the robustness of the scenario found for the $T=0$ RFIM with the standard single-spin-flip dynamics, it has been proposed to consider the nonequilibrium response of the RFIM still at $T=0$, but with dynamics involving possible cooperative flips of several spins\cite{VRT2005}. In such dynamics, the system can effectively overcome energy barriers that would block it in a one-spin-flip dynamics. Even if thermally activated processes are absent, the metastable states visited by the system when the applied field is ramped up or down are a priori different from those reached in the one-spin-flip dynamics. A simulation of the $T=0$ RFIM with 2-spin-flip dynamics on a cubic lattice showed that the change of dynamics yields a significant reduction of the width of the hysteresis loop (or, in magnetic language, of the coercivity), but that the out-of-equilibrium critical behavior is still present and appears, within statistical uncertainty, to belong to the same universality c!
 lass as that observed when using the standard one-spin-flip dynamics\cite{VRT2005}.

In this article, we study the hysteresis behavior of the RFIM on a Bethe lattice with a 2-spin-flip dynamics at zero temperature. We consider a gaussian distribution of the random fields. Our goal is to obtain an exact solution for the main hysteresis loop as a function of disorder strength, and to compare the results with those of the 1-spin-flip dynamics already derived\cite{S1996,DSS1997}. A question that was found intriguing in  the latter case is the absence of out-of-equilibrium transition for a Bethe lattice with connectivity z=3\cite{DSS1997}. A transition is found for $z\geq4$ and no transition occurs when $z=2$ (the one-dimensional case). The fact that a qualitative change takes place for $z>3$ and not, as found in equilibrium, for $z>2$ is, to the least, unusual. It casts a doubt on the conjecture that the out-of-equilibrium and equilibrium critical points belong to the same universality class and are controlled by the same fixed point \cite{SDP2004,PV2004,A2005}.!
  We thus focus on the Bethe lattice with connectivity $z=3$ and we show that an out-of-equilibrium transition does occur in this case with the 2-spin-flip dynamics.

The rest of the paper is organized as follows. In section II we present the model and the dynamics, contrasting the 2-spin-flip dynamics with the standard 1-spin-flip one and reviewing the main properties of the 2-spin-flip dynamics. The next section is devoted to deriving the self-consistent equations that allow to calculate the main hysteresis loop on a Bethe lattice. In section IV, we consider the one-dimensional chain ($z=2$) and in section V, the $z=3$ case. In both cases, we compare with the results obtained with the 1-spin-flip dynamics as well as with the equilibrium (ground state) behavior. For the connectivity $z=3$, we show that, in contrast with the 1-spin-flip dynamics, the 2-spin-flip dynamics leads to a disorder-induced, out-of-equilibrium transition. Finally, in the last sections, we discuss possible extensions of the approach  to the calculation of other observables and to more general k-spin-flip dynamics, with an emphasis on the influence of the cooperativ!
 ity of the (local) dynamics on the hysteresis properties, and we conclude.

\section{Model and dynamics}
\label{model}

We study the RFIM on a lattice of connectivity z, with Ising spin variables $S_{i}=\pm1$ ferromagnetically interacting with their neighbors and subject to both an intrinsic random field and an applied uniform field H. The Hamiltonian is given by
\begin{equation}
H=-J\sum_{<ij>}S_{i}S_{j}-\sum_{i}(h_{i}+H)S_{i},
\end{equation}
where the sum $<ij>$ is over all distinct ``bonds"  of the lattice and the random fields $h_{i}$ on each site  are independently drawn from a gaussian probability distribution of zero mean and variance equal to $\sigma$, $\rho(h) =\exp(-h^2/2 \sigma^2)/\sqrt{2\pi}\sigma$ .

We consider the limit of zero temperature and the evolution of the system as it is (adiabatically) driven by the external magnetic field $H$. At $T=0$, metastability results from the absence of full equilibration and the presence of several distinct configurations in which the system can be trapped. It is customary to define such configurations as the local energy minima, namely the states that minimize the Hamiltonian in Eq. (1) (due to the presence of the applied magnetic field, the energy is actually a magnetic enthalpy, but in the following we shall keep a loose usage of the term "energy"), and to envisage the phenomenon within the topographic picture of a rugged energy landscape. For Ising spins,  the energy levels are discretized and the topographic view requires some adjustments. A useful characterization of the minima consists in sorting them according to the number of spins that must be flipped to yield a lower-energy state. More precisely, one can define a $k$-stab!
 le minimum as a configuration whose energy cannot be decreased by flipping any subset of $k$, or less, spins\cite{BM2000}. The usual minima are then 1-stable configurations, which are stable with respect to the flip of any single spin in the system. The ground state of the system on the other hand corresponds to a configuration that is stable with respect to any arbitrary large number of (simultaneous) spin flips, \textit{i. e.}, to the limit $k\rightarrow \infty$\cite{BM2000,MP2003}.

For classical spins with no intrinsic dynamics, knowledge of the energy landscape is not enough if one does not provide a dynamical rule to explore the landscape. For the problem of interest here, most studies have considered the standard 1-spin-flip dynamics. Under such dynamics, the system evolves along irreversible paths among local energy minima which by construction are 1-stable for the relevant range of applied field $H$. At each site $i$, the spin $S_{i}$ aligns with the net local field $f_{i}$, with
\begin{equation}
f_{i}=J\sum_{j(i)}S_{j}+h_{i}+H,
\end{equation}
where the sum is over all connected neighbors of site $i$.

A natural extension of this dynamics consists in allowing cooperative flips of pairs of spins. The system then jumps from one state to another (via an avalanche) whenever the former state becomes unstable to the flip of any single spin \textit{or} to the simultaneous flip of any single pair of spins. Evolution now proceeds among 2-stable local minima. New features with respect to the 1-spin-flip dynamics are only introduced when the flipping pair is made up by two neighboring spins, and the local energy to be minimized is then
\begin{equation}
H_{ij}=-f'_{i}S_{i}-f'_{j}S_{j}-JS_{i}S_{j}
\end{equation}
where
\begin{equation}
f'_{i}=J\sum_{k(i)\neq j}S_{k}+h_{i}+H
\end{equation}
is the local field experienced by $S_{i}$ without the influence of the neighbor $S_{j}$ (the sum in the above expression is over all neighbors of $i$ except $j$). The transition of the pair $(ij)$ from one state to another under the 2-spin-flip dynamical rule is graphically illustrated in Figure 1. One can think the dynamics as made of both single-spin-flips and "irreducibly cooperative" 2-spin flips. The former occur whenever the net local field on $S_{i}$, $f_{i}=f'_{i}+JS_{i}$, or $S_{j}$, $f_{j}=f'_{j}+JS_{i}$, changes sign. This corresponds to the transitions $--\longleftrightarrow -+$ , $--\longleftrightarrow +-$ and $-+\longleftrightarrow ++$ , $+-\longleftrightarrow ++$ in the diagram. An irreducibly cooperative 2-spin-flip involves a pair of neighboring spins which are of the same sign ($--$ or $++$) and cannot flip individually, i.e., without the simultaneous flip of the neighbor. This takes place whenever the net field on the pair of aligned spins, $f_{ij}=f'_{i}+!
 f'_{j}$, changes sign in the central region of the diagram: $-J<f'_{i}<J$ (condition for forbidding spin $S_{i}$ to flip individually) and $-J<f'_{j}<J$ (condition for forbidding spin $S_{j}$ to flip individually). The applied magnetic field is varied until a first pair of spins become marginally stable. This pair can be represented by a point on the diagram in Figure 1 and the further evolution of the pair is dictated by the border that is first attained (notice that the axes $f'_{i}$ and $f'_{j}$ depend linearly on $H$ so that the displacement of the point is always along the $(1,1)$ diagonal).
\begin{center}
\begin{figure}[th]
\includegraphics[width=7cm,clip]{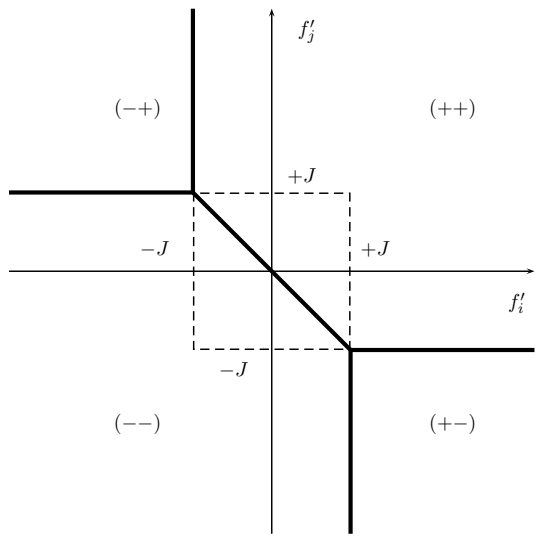}
\caption{\label{fig1} Stability diagram for a pair of nearest-neighbor spins $i,j$ under the 2-spin-flip dynamical rule. The axes are the local fields defined in Eq. (4).}
\end{figure}
\end{center}

In Ref.\cite{VRT2005}, it was shown that, due to the ferromagnetic nature of the interactions, the 2-spin-flip dynamics shares with the 1-spin-flip one important properties, known as the no-passing rule, the abelian property, and the existence of return-point memory. As a result, (i)  the final 2-stable configuration reached through an avalanche is independent of the order in which the unstable spins have been ``relaxed", i. e. flipped to minimize the energy), (ii) the state reached at a given applied field is independent of the details of the previous history and only depends on the state in which the field was last reversed, provided the field is adiabatically varied, and (iii) in the relaxation process between two metastable states after a change of the applied field, no spins flip more than once.

In the following we focus on the main hysteresis loop obtained by starting from a satured configuration with all spins pointing in the same direction in a very large (positive or negative) applied field. Because of the symmetry of the model, the upper branch of the loop followed by the system when decreasing the applied field from $+\infty$ to $- \infty$ is easily deduced from the lower branch obtained by increasing the field from $-\infty$ to $+\infty$, and we therefore only consider the latter. We search for an exact solution of the problem on a Bethe lattice of connectivity $z$.

\section{Self-consistent equations on a Bethe lattice}
\label{formalism}

As is standard in statistical physics, we consider a Bethe lattice as the "deep interior" of a Cayley tree in the limit where the number of sites in both go to infinity, the size of the whole tree going to infinity faster than that of the chosen interior piece\cite{MP2003}. The tree structure allows to write recursion relations and the above trick allows to close the relations by removing the effect of the boundaries. A detailed analysis of the hysteresis behavior of the $T=0$ RFIM on a Bethe lattice with the 1-spin-flip dynamics has already been provided\cite{S1996,DSS1997,A2005,S2001,SSD2000,CGZ2002, IOV2005}. Extension to more general dynamics is however nontrivial and we detail here the main aspects of the formalism.

We first choose at random a site in the deep interior of the Cayley tree of connectivity $z$ and take it as the origin. From this origin, one can label the ``descendant" sites according to their level in the tree. A spin at level $n$ will then have one parent spin at level $(n-1)$ and $z-1$ descendant spins at level $(n+1)$: see Figure 2. Our goal is to compute the average magnetization as a function of the magnetic field $H$ in the lower branch of the hysteresis loop. By using the properties of the dynamics, it is then sufficient to start from a configuration where all spins are down and to follow the relaxation of the spins starting from the outer part of the lattice and progressing inward, toward the origin, until a 2-stable configuration at the given value $H$ of the field is reached.
\begin{center}
\begin{figure}[th]
\includegraphics[width=7cm,clip]{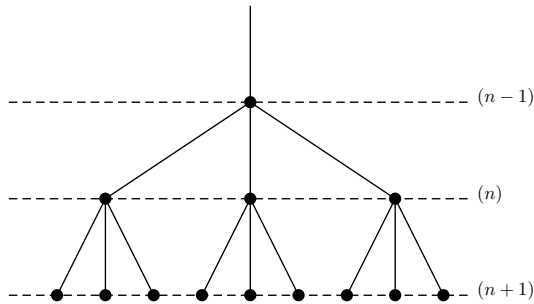}
\caption{\label{fig2} llustration of the branching characteristics of the Cayley tree for $z=4$.}
\end{figure}
\end{center}

In the 1-spin-flip dynamics, the central quantity is the conditional probability that a spin at level $n$ is flipped in the relaxation process, \textit{given} that its parent at level $(n-1)$ is kept down. This conditional probability is a function of the external field only (it is already an average with respect to the random fields) and it obeys a recursion equation whose fixed point describes the typical behavior on the Bethe lattice. Deriving the counterpart of this quantity for the 2-spin-flip dynamics is a little tricky. Indeed, cooperative flips of pairs of spins couple the levels in a more intricate way. The appropriate quantity is actually the conditional probability that a spin at level $n$ is in state $S$ (with $S=\pm1$, where as usual ``$+/-$'' and ``up/down'' are used interchangeably) \textit{given} that: (i) its local random field is $h$, (ii) all spins (and pairs of spins) in the branch are relaxed \textit{with the constraint} that the parent spin at level $(n!
 -1)$ is kept down, and (iii) the $z-1$ descendant spins at the level $(n+1)$ are in states $S_{1}, S_{2},.., S_{z-1}$. The probability is now a function of the sum of the random and the uniform fields, $x=h+H$, and it is easy to show that it depends on the state of the descendant spins at level $(n+1)$ only through the sum $\tilde{S}=\sum_{i=1}^{z-1}S_{i}$. We denote this conditional probability $p^{(n)}_{\tilde{S}S }(x)$.

To write the recursion relation for this probability, one needs to introduce 2-spin-flip transition rates involving the change of the configuration formed by a spin at level $n$ and its $z-1$ direct descendants, \textit{given} the state of the parent at level $(n-1)$ \textit{and} the state of all the $(z-1)^2$ descendants at level $(n+2)$. Let us  denote 
\begin{equation}
W_{\{ \tilde{\bf S}_{(n+2)}|({\bf S}'_{(n+1)},S'_{(n)}) \rightarrow ({\bf S}_{(n+1)},S_{(n)})|S_{(n-1)}\} }({\bf x}_{(n+1)},x_{(n)})
\end{equation}
the transition rate for going from the configuration $({\bf S}'_{(n+1)},S'_{(n)}) $, where ${\bf S}'_{(n+1)}\equiv \{S'_1,..., S'_{z-1}\}_{(n+1)}$ to the configuration $({\bf S}_{(n+1)},S_{(n)}) $, where ${\bf S}_{(n+1)}\equiv \{S_1,..., S_{z-1}\}_{(n+1)}$, \textit{given} that the parent  at level $(n-1)$ is in state $S_{(n-1)}$ and the descendants at level $(n+2)$ are in states $\tilde{\bf S}_{(n+2)}\equiv \{\tilde{S}_1,...,\tilde{S}_{z-1}\}_{(n+2)}$ (again, it is easy to show that the rate only depends on the sum of the states of the $z-1$ descendants merging at the common root at level $(n+1)$, e.g., $\tilde{S}_{1}=\sum_{i=1}^{z-1}S_{i\rightarrow 1}$); the rate depends on the local fields at level $n$, $x_{(n)}$, and at level $(n+1)$, ${\bf x}_{(n+1)}\equiv \{x_1,...,x_{z-1}\}_{n+1}$. We shall only be interested in the situation where $S'_{(n)}=-1$ and $S_{(n-1)}=-1$. A graphical representation is given in Figure 3. The transition rate can be written by combining stabilit!
 y diagrams similar to that in Figure 1 for all the pairs of spins involved in the change of state. An explicit expression will be provided below for the case $z=2$.
\begin{center}
\begin{figure}[th]
\includegraphics[width=4cm,clip]{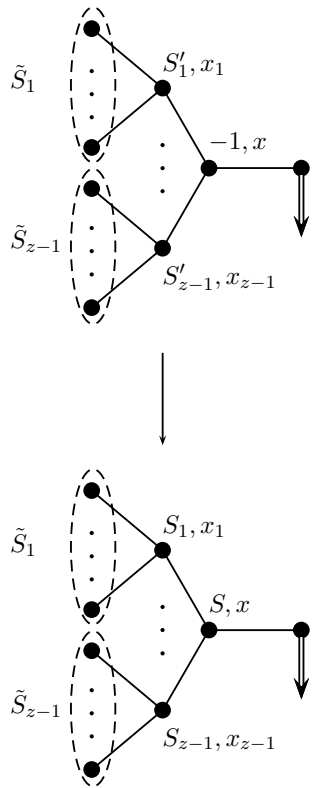}
\caption{\label{fig3} Schematic illustration of the transition rate $W_{\{ \tilde{\bf S}_{(n+2)}|({\bf S}'_{(n+1)},S'_{(n)}) \rightarrow ({\bf S}_{(n+1)},S_{(n)})|S_{(n-1)}\} }({\bf x}_{(n+1)},x_{(n)})$ involved in the recursion equations for  the conditional probabilities (see Eq. (5)). The configuration formed by a spin (initially down) and its $z-1$ descendants changes via both single-spin flips and irreducibly cooperative $2$-spin flips, given that the parent (the rightmost spin) is kept down and that the descendants on the left are in states 
$(\tilde{S}_1,\ldots \tilde{S}_{z-1})$.}
\end{figure}
\end{center}

The recursion relation for the conditional probability $p^{(n)}_{S \tilde{S}}(x)$ can now be derived by considering all ways to generate the chosen configuration at level $n$ from configurations on the $z-1$ branches at level $(n+1)$. One obtains
\begin{align}
p^{(n)}_{\tilde{S}S}(x)=Tr_{\tilde{\bf S}, {\bf  S'}} C_{z-1}^{(\tilde{S}_{1}+z-1)/2}\ldots & C_{z-1}^{(\tilde{S}_{z-1}+z-1)/2}\int\!\!\delta  x_{1}\ldots\delta  x_{z-1} p^{(n+1)}_{\tilde{S}_{1}S'_{1}}(x_{1})\ldots p^{(n+1)}_{\tilde{S}_{z-1}S'_{z-1}}(x_{z-1})\nonumber\\ 
&\times W_{\left\lbrace \tilde{\bf S}_{(n+2)}|({\bf S}'_{(n+1)},-1) \rightarrow ({\bf S}_{(n+1)},S|-1\right\rbrace }({\bf x}_{(n+1)},x)
\end{align}
where we have used the short-cut notation $\delta x \equiv  \rho(x-H)dx$ and $\{S_{1} \ldots S_{z-1}\}$ is any configuration such that $S_{1}+\ldots+S_{z-1}=\tilde{S}$; as already stressed, it is indeed easy to prove that the conditional probability is the same for all such configurations: $p^{(n)}_{\tilde{S}S}(x)\equiv p^{(n)}_{(S_{1}\ldots S_{z-1})S}(x)$. The sums on the $\tilde{S}_{i}$'s are restricted to the values $2r-(z-1)$ with $r$ an integer between $0$ and $z-1$ (the combinatorial factors then count the number of different configurations of $z-1$ spins with a sum equal to $2r-(z-1)$). The recursion equation can be closed by making the ``homogeneous" assumption that all sites in the deep interior are equivalent, provided they have the same random field. This amounts to consider the fixed point of the recursion, which is solution of the self-consistent equation obtained by dropping the superscripts $n$ and $n+1$ in the above equation.

One more step is still required to compute the average magnetization. This latter is simply related to the average probability that the central spin (at the randomly chosen origin of the Bethe lattice) is up:
\begin{equation}
m(H)=2\bar{P}(+1;H)-1,
\end{equation}
where the overbar denotes an average over the random field distribution, $\bar{P}(+1;H)=\int \delta x P(+1;x,H)$. The probability that the central spin is up is obtained by studying the merging of the $z$ branches of the tree at the origin. The state of each branch \textit{given} that the central spin is kept down is independent from that of the other branches and is characterized by the conditional probability $p_{\tilde{S}S}(x)$. Full relaxation implies to relax the constraint on the central spin and let the system evolve according to the 2-spin-flip dynamical rule. This evolution is described with the help of transition rates $\tilde W$ that are given by an extension of those introduced above: the transition now involves configurations formed by the central spin and its $z$ neighbors, given the states of the direct descendants of those neighbors, see Figure 4. One  then obtains
\begin{equation}
\begin{split}
\bar{P}(+1;H)=\int\!\!\delta x\sum_{\tilde{S}_1...\tilde{S}_z}\sum_{S'_{1}...S'_{z}}\sum_{S_{1}...S_{z}}\int\!\!\delta x_{1} \cdots \delta x_{z}\; p_{\tilde{S}_1S_{1}}(x_{1}) \cdots p_{\tilde{S}_zS_{z}}(x_{z})\\ \times \tilde W_{\{\tilde{S}_{1}\ldots\tilde{S}_{z}|(S'_1\ldots S'_z,-1)\rightarrow(S_{1}\ldots S_{z},+1)\}}(x_{1}\ldots x_{z},x),
\end{split}
\end{equation}
where  $\tilde W$ is the transition rate defined just above; to simplify the notations we have  included, here and in the rest of the paper, the combinatorial factor  $C_{z-1}^{(\tilde{S}+z-1)/2}$ in a redefinition of the conditional probability $p_{\tilde{S}S}(x)$.

Having given the general formalism to calculate the main hysteresis loop on a Bethe lattice under a 2-spin-flip dynamics, we now specialize the study to the cases $z=2$ and $z=3$.
\begin{center}
\begin{figure}[th]
\includegraphics[width=4cm,clip]{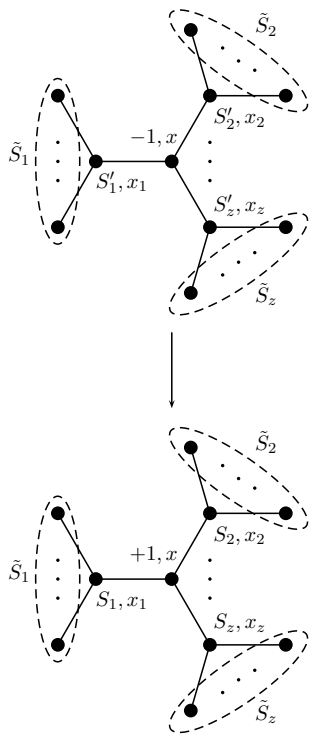}
\caption{\label{fig4} Schematic illustration  of the transition rate $W_{\{\tilde{S}_{1}\ldots\tilde{S}_{z}|(S'_1\ldots S'_z,-1)\rightarrow(S_{1}\ldots S_{z},+1)\}}(x_{1}\ldots x_{z},x)$ involved in the expression of the magnetization (see Eq. (8)). The central spin flips up in a change of configuration that involves its $z$ neighbors, given the states of the descendants $(\tilde{S}_1,\ldots \tilde{S}_{z})$. The change of configuration involves both single-spin flips and irreducibly cooperative $2$-spin flips.}
\end{figure}
\end{center}

\section{Exact solution for the one-dimensional chain}
\label{one-dimensional}
The case of the one-dimensional chain ($z=2$) is of course the easiest to handle. The self-consistent equation obtained from Eq. (6) is linear in the conditional probability, but in contrast with what occurs for the 1-spin-flip dynamics, the probability is a nontrivial function of the local field and Eq. (6) has the form of an integral equation:
\begin{equation}
p_{\tilde{S}S}(x)=\sum_{\tilde{S}_{1}}\sum_{S_{1}} \int\!\!\delta x_{1}\;p_{\tilde{S}_{1}S_{1}}(x_{1})
\; W_{\{\tilde{S}_{1}|(S_{1},-1)\rightarrow(\tilde{S},S)|-1\}}(x_{1},x).
\end{equation}

 We first specify the explicit form of the transition rates that appear in this equation. We consider the transition for a pair of neighbors \textit{given} that the descendant is in state $\tilde{S}$ and that the parent is down. Recalling the relation between the local fields $f'_{i}$ introduced in section II and the local field $x_{i}$ used in section III (namely, $f'_{i}$ is equal to $x_{i}$ plus the field due to the interaction with the neighbor that does not belong to the pair under consideration) and that the external field $H$ increases, we can adapt the stability diagram of Figure 1  to describe the different terms and sectors composing the transition rates. The resulting expressions are given below:
\begin{align}
W_{\{\tilde{S}_{1}|(--)\rightarrow(++)|-\}}(x_{1},x)&=
\theta(0<x<2J)\theta(-J(\tilde{S}_{1}-1)-x<x_1<-J(\tilde{S}_{1}-1))+\nonumber\\
&\theta(x>2J)\theta( -J(\tilde{S}_{1}+1)<x_1<-J(\tilde{S}_{1}-1) )\nonumber\\
W_{\{\tilde{S}_{1}|(--)\rightarrow(-+)|-\}}(x_{1},x)&=
\theta(x>2J)\theta(x_1<-J(\tilde{S}_{1}+1))\nonumber\\
W_{\{\tilde{S}_{1}|(--)\rightarrow(+-)|-\}}(x_{1},x)&=0\nonumber\\
W_{\{\tilde{S}_{1}|(--)\rightarrow(--)|-\}}(x_{1},x)&=
\theta(x<0)\theta(x_1<-J(\tilde{S}_{1}-1))+
\theta(0<x<2J)\theta(x_1<-J(\tilde{S}_{1}-1)-x)\nonumber\\
W_{\{\tilde{S}_{1}|(+-)\rightarrow(++)|-\}}(x_{1},x)&=
\theta(x>0)\theta(x_1>-J(\tilde{S}_{1}-1))\nonumber\\
W_{\{\tilde{S}_{1}|(+-)\rightarrow(-+)|-\}}(x_{1},x)&=0\nonumber\\
W_{\{\tilde{S}_{1}|(+-)\rightarrow(++)|-\}}(x_{1},x)&=
\theta(x<0)\theta(x_1>-J(\tilde{S}_{1}-1))\nonumber\\
W_{\{\tilde{S}_{1}|(+-)\rightarrow(--)|-\}}(x_{1},x)&=0
\end{align}
where $\theta(...)$ is the characteristic function of the domain of local field indicated by the argument: it is $1$ in the domain and $0$ outside; with this definition, $\theta(x>0)$ is the usual Heaviside function $H(x)$, $\theta(0<x<2J)$ is equal to $H(x)-H(x-2J)$, and so on.

By inserting the above expressions into Eq. (6), one finds that the conditional probabilities satisfy the following equations:
\begin{align}
p_{++}(x)&= \theta(0<x<2J)\left[ \sum_{\tilde{S}_1}\int_{-J(\tilde{S}_1-1)}^{+\infty} \!\!\delta x_1 p_{\tilde{S}_1+}(x_1)+\sum_{\tilde{S}_1}\int_{-J(\tilde{S}_1-1)-x}^{-J(\tilde{S}_1-1)} \!\!\delta x_1 p_{\tilde{S}_1-}(x_1)  \right] \nonumber\\
&            +\theta(x>2J) \left[\sum_{\tilde{S}_1}\int_{-J(\tilde{S}_1-1)}^{+\infty} \!\!\delta x_1p_{\tilde{S}_1+}(x_1)+\sum_{\tilde{S}_1}\int_{-J(\tilde{S}_1+1)}^{-J(\tilde{S}_1-1)} \!\!\delta x_1 p_{\tilde{S}_1-}(x_1) \right]\nonumber\\
p_{-+}(x)&= \theta(x>2J)\left[ \sum_{\tilde{S}_1}\int_{-\infty}^{-J(\tilde{S}_1+1) } \!\!\delta x_1 
                                 p_{\tilde{S}_1-}(x_1)\right]\nonumber\\
p_{+-}(x)&= \theta(x<0)\left[ \sum_{\tilde{S}_1}\int_{-J(\tilde{S}_1-1)}^{+\infty} \!\!\delta x_1 
                                 p_{\tilde{S}_1+}(x_1)\right]\nonumber\\
p_{--}(x)&= \theta(x<0)\left[ \sum_{\tilde{S}_1}\int_{-\infty}^{-J(\tilde{S}_1-1) } \!\!\delta x_1 
                                 p_{\tilde{S}_1-}(x_1)\right] 
          +\theta(0<x<2J)\left[ \sum_{\tilde{S}_1}\int_{-\infty}^{-J(\tilde{S}_1-1)-x} \!\!\delta x_1 
                                 p_{\tilde{S}_1-}(x_1)\right] 
\end{align}

Notice that the dependence on $x$ comes through both the characteristic functions and, in the case of $p_{++}$ and $p_{--}$, through an additional dependence in the region $0<x<2J$ (see the bounds of the integrals).

It is convenient to introduce the disorder-averaged values of the conditional probabilities,
\begin{equation}
\bar{p}_{S \tilde{S}}=\int \delta x\; p_{S \tilde{S}}(x).
\end{equation}
\begin{center}
\begin{figure}[th]
\includegraphics[width=7cm,clip]{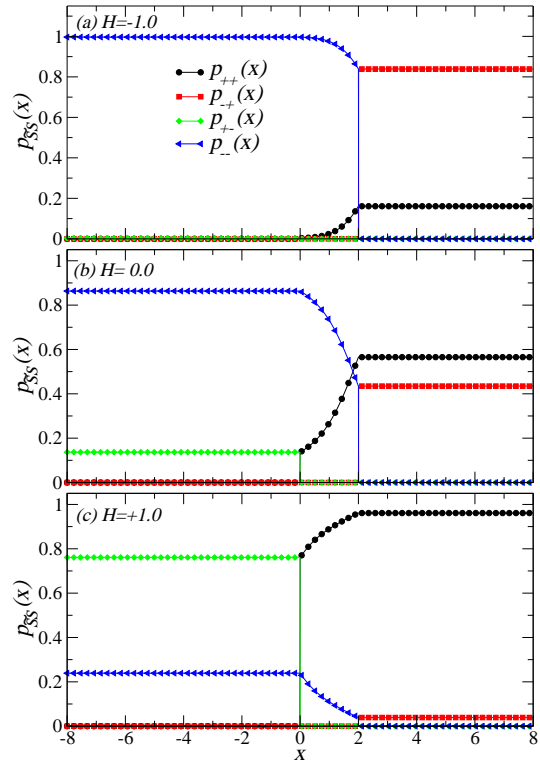}
\caption{\label{fig5}(Color on line) Conditional probabilities $p_{\tilde{S}S}(x)$ as a function of the local field $x$ for the 1-dimensional chain ($z=2$) and for $\sigma= 1.0$. Different values of the applied field are considered (here and in all figures the fields and the disorder strength are in units $J\equiv 1$). Note the continuity properties discussed in the text. }
\end{figure}
\end{center}

These quantities can be computed directly in computer simulations, as will be discussed below. As detailed in Appendix A, the solution to the set of coupled linear integral equations formed by  Eqs. (11) can be found by reexpressing all the terms needed in the expression of the conditional probabilities with the help of the disorder-averaged values and of a function $q(x)$ that is nonzero only in the domain $0<x<2J$ (where it is equal to $p_{--}(x)$, by definition). After some algebra, one finds
\begin{align}
\bar{p}_{++}&=p_0p_1(1+p_2-p_1)+p_1-p_0-[1+p_0(p_2-p_1)]\int_0^{2J}\delta xq(x)\nonumber\\
\bar{p}_{+-}&=p_1(1-p_1)-(1-p_1)\int_0^{2J}\delta xq(x)\nonumber\\
\bar{p}_{-+}&=p_0[1-p_1(1+p_2-p_1)]+p_0(p_2-p_1)\int_0^{2J}\delta xq(x)\nonumber\\
\bar{p}_{--}&=(1-p_1)^2+(2-p_1)\int_0^{2J}\delta xq(x)
\end{align}
where the function $q(x)$ satisfies an inhomogeneous integral equation,
\begin{equation}
q(x)-\int_0^{2J-x}\delta x_1q(x_1)=1-p_1(1+\int_{-x}^{0}\delta x_1)+\big(\int_{-x}^{0}\delta x_1\big)\int_0^{2J}\delta x_1q(x_1)
\end{equation}
for $0<x<2J$. In the above expressions, we have introduced the standard probabilities that play a key role for the 1-spin-flip dynamics, $p_{n}(H)=\int_{(z-2n)J}^{+\infty} \delta x\equiv \int_{(z-2n)J-H}^{+\infty} dh\rho(h)$, with $n=0,1$, or $2$: $p_{n}(H)$ is the probability that a spin with $n$ neighbors up and $z-n$ neighbors down flips up in an applied field H via the 1-spin-flip mechanism. The probabilities $p_{\tilde{S}S}(x)$ satisfy the normalization property
\begin{equation}
p_{++}(x)+p_{+-}(x)+p_{-+}(x)+p_{--}(x)=1,
\end{equation}
for any $x$. (As a result the normalization condition is also satisfied by the disorder-averaged probabilities, as can be verified by using Eqs. (13).) It is also easy to check that the functions $p_{\tilde{S}S}(x)$ satisfy continuity relations: for instance, $p_{++}(0^{+})=p_{+-}(0^{-})$ and $p_{--}(2J^{-})=p_{-+}(2J^{+})$. Finally, one can recover the results of the 1-spin-flip dynamics\cite{S1996} by setting $q(x)=(1-p_1)/(1-p_1+p_0)$ over the whole interval $\left[ 0,2J\right] $. 
\begin{center}
\begin{figure}[th]
\includegraphics[width=7cm,clip]{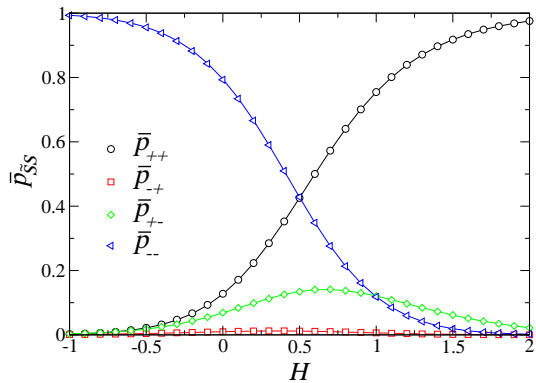}
\caption{\label{fig6} (Color on line) Average conditional probabilities as a function of the applied field $H$ for $z=2$ and $\sigma=1.0$. The lines correspond to the exact solution of Eqs. (13,14) and the symbols to the simulation results  discussed in section VI ($N_{sites}=10^4$, $N_{realizations}=10^5$).}
\end{figure}
\end{center}

We have numerically obtained the solution of Eq. (14) (the solution can be computed with any arbitrary precision), from which we have calculated the conditional probabilities and their average. The resuts are illustrated in Figures 5 and 6. The continuity relations are clearly seen on Figure 5. One moreover finds that the average probabilities are smooth (continuously differentiable) functions of the external field for all values of the disorder strength $\sigma$.

Going from the conditional probabilities to the average magnetization needs the additional step provided by Eq. (8). The  expressions of the transition rates are now a little more complicated than those in Eqs. (10), since one must consider the different transition possibilities for a configuration of 3 spins (see Figure 4 with $z=2$). The explicit formulae are given in Appendix A 

The resulting magnetization curve $m(H)$ corresponding to the lower branch of the hysteresis loop is shown in Figure 7 for several values of the disorder strength $\sigma$. Again, the curves are always smooth, irrespective of the value of $\sigma$. As expected, no out-of-equilibrium transition takes place in the 2-spin-flip dynamics of the $T=0$ RFIM in one dimension. This will be discussed in more detail below.
\begin{center}
\begin{figure}[th]
\includegraphics[width=7cm,clip]{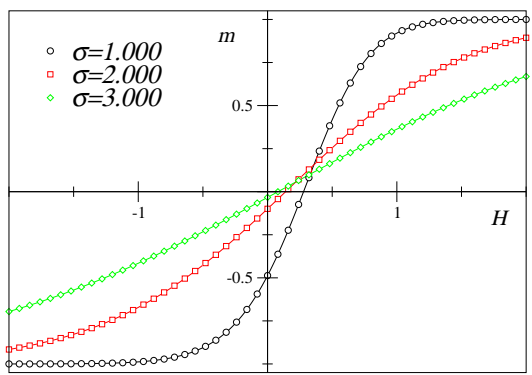}
\caption{\label{fig7} (Color on line) Lower branch of the hysteresis loop for several values of the disorder ($z=2$). The lines correspond to the exact solution and the symbols to the simulation results ($N_{sites}=10^4$, $N_{realizations}=10^2$). }
\end{figure}
\end{center}

\section{Bethe lattice with connectivity $z=3$}
\label{$z=3$}

As mentioned in the Introduction, the case $z=3$ is especially interesting. Whereas a disorder-induced transition (critical point) is found in the ground state, no such transition occurs in the nonequilibrium response with 1-spin-flip dynamics. The calculation of the hysteresis loop under 2-spin-flip dynamics is, however, considerably more involved than in the one-dimensional case.
\begin{center}
\begin{figure}[th]
\includegraphics[width=7cm,clip]{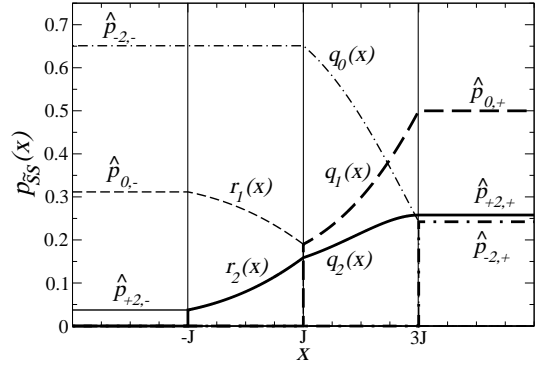}
\caption{\label{fig8} Conditional probabilities $p_{{\tilde S}S}(x)$ as a function of the local field $x$ for $z=3$ ($H= 0.4$ and $\sigma=1.1$). The sum of the probabilities is equal to $1$ in all intervals. The plateaus correspond to the constants ${\hat p}_{{\tilde S}S}$  which, as the functions in the intervals $[-J,J]$ and $[J,3J]$, are defined in Eqs. (16).}
\end{figure}
\end{center}

Specializing the self-consistent equations for the conditional probabilities, Eqs. (6), to the case $z=3$ leads to nonlinear coupled integral equations. The explicit form of the transition rates must now be derived by considering the two pairs formed by a spin and its two descendants and by combining two diagrams similar to that of Figure 1 to form a three-dimensional diagram with three local fields as axes.  The number of different conditional probabilities, $p_{\tilde{S},S}(x)$, is six (corresponding to $\tilde{S}=-2,0,+2$ and $S=+,-$), because the dependence on the states of the two descendant spins comes only through their sum. The dependence of these conditional probabilities on the local field $x$ is piecewise, as in the $z=2$ case. However, contrary to the $z=2$ case, several nontrivial functions of $x$ in the intervals $\left[-J,J \right] $ and $\left[J,3J \right]$ must be introduced instead of only one . As discussed in more detail in Appendix B, the general form of!
  the conditional probabilities is as follows:
\begin{align}
p_{+2,+}(x)&=\theta(-J<x<J) r_2(x) +\theta(J<x<3J) q_2(x)+\theta(x>3J)\hat{p}_{+2,+}\nonumber\\
p_{0,+}(x)&=\theta(J<x<3J) q_1(x) +\theta(x>3J) \hat{p}_{0,+}\nonumber\\
p_{-2,+}(x)&=\theta(x>3J)\hat{p}_{-2,+}\nonumber\\ 
p_{+2,-}(x)&=\theta(x<-J)\hat{p}_{+2,-}\nonumber\\
p_{0,-}(x)&=\theta(x<-J) \hat{p}_{0,-} +\theta(-J<x<J) r_1(x)\nonumber\\
p_{-2,-}(x)&=\theta(x<J) \hat{p}_{-2,-} +\theta(-J<x<3J) q_0(x)
\end{align}
where the expressions for the $6$ constants, $\hat{p}_{\tilde{\alpha},\alpha}$, and the $5$ functions, $q_0(x), q_1(x),q_2(x)$ and $r_1(x)$ and $r_2(x)$, are given in Appendix B. The continuity relations are illustrated in Figure 8.  The conditional probabilities satisfy a normalization property, which due to their piecewise form can be written as
\begin{align}
&\hat{p}_{+2,+}+\hat{p}_{0,+}+\hat{p}_{-2,+}=1\nonumber\\
&\hat{p}_{-2,-}+\hat{p}_{0,-}+\hat{p}_{+2,-}=1\nonumber\\
&q_0(x)+q_1(x)+q_2(x)=1\nonumber\\
&\hat{p}_{-2,-}+r_1(x)+r_2(x)=1
\end{align}
\begin{center}
\begin{figure}[th]
\includegraphics[width=9cm,clip]{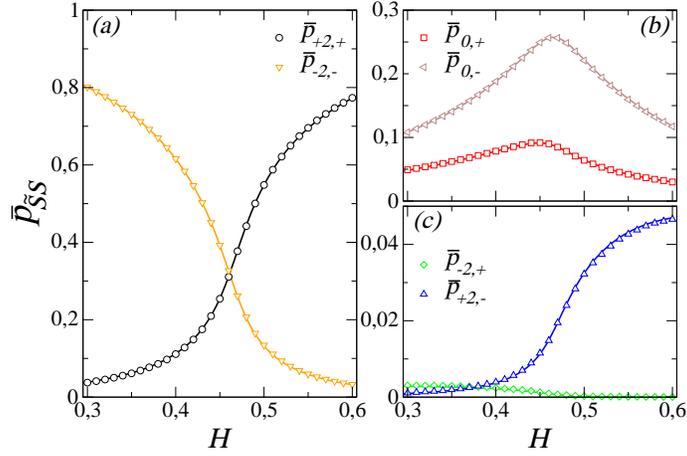}
\caption{\label{fig9} (Color on line) Average conditional probabilities as a function of the applied field for $z=3$  (strong-disorder regime, $\sigma =1.1$): (a) $\bar{p}_{+2,+}$ and $\bar{p}_{-2,-}$, (b) $\bar{p}_{0,+}$ and $\bar{p}_{0,-}$, (c) $\bar{p}_{-2,+}$ and $\bar{p}_{+2,-}$ (note the changes of scale).
All probabilities are continuous functions of $H$; $\bar{p}_{+2,+}$, $\bar{p}_{-2,-}$,  $\bar{p}_{-2,+}$ and $\bar{p}_{+2,-}$ vary monotonically whereas  $\bar{p}_{0,+}$ and 
$\bar{p}_{0,-}$ go through a maximum. The lines represent the solution of the exact self-consistent equations and the symbols the simulation results ($N_{sites}=10^5$, $N_{realizations}=2.10^5$).}
\end{figure}
\end{center}

As a result, one has to consider only $4$ constants and $3$ functions. The latter are solutions of coupled inhomogeneous nonlinear integral equations (more details are provided in Appendix B). We have solved the set of equations (B1-B2) by an iterative method, starting with $\hat{p}_{-2,-}=1$, $q_0(x)=1$,  $\hat{p}_{-2,+}=1$, and all other functions equal to zero as initial conditions. The ``new'' values of the $p_{\tilde{S}S}(x)$'s  are then obtained by computing the integrals in  the right-hand side of Eqs. (B1-B2) with the ``old'' values, using  Simpson's rule. Convergence is assumed when  max$\vert p_{\tilde{S}S}^{new}-p_{\tilde{S}S}^{old}\vert <10^{-4}$.
It is important to notice that, for a given external field, these equations 
may have several solutions (as discussed just below). When this happens, we find that only two solutions are stable, in the sense that they can be reached by the iteration algorithm. In addition to the solution obtained with the above mentioned initial conditions, the other one can be found by starting with 
$\hat{p}_{+2,-}=1$, $r_2(x)=1$, $q_2(x)=1$, $\hat{p}_{+2,+}=1$, and the 
other functions  equal to zero.
\begin{center}
\begin{figure}[th]
\includegraphics[width=9cm,clip]{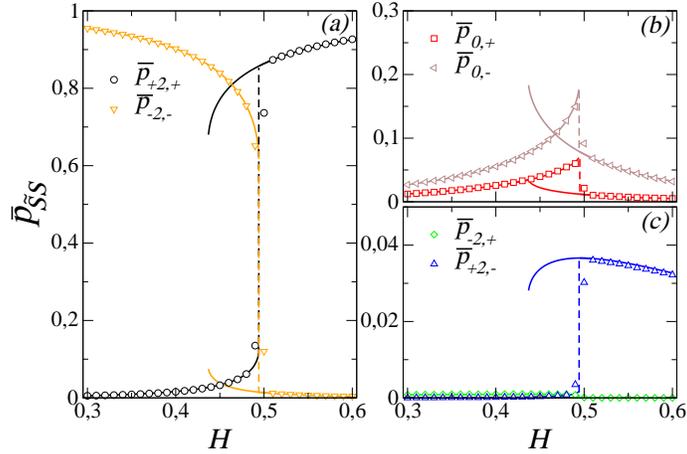}
\caption{\label{fig10} (Color on line) Average conditional probabilities as a function of the applied field  for $z=3$ (weak-disorder regime, $\sigma=0.9$): (a) $\bar{p}_{+2,+}$ and $\bar{p}_{-2,-}$, (b) $\bar{p}_{0,+}$ and 
$\bar{p}_{0,-}$,  (c) $\bar{p}_{-2,+}$ and $\bar{p}_{-2,-}$ (note the changes of scale). The probabilities show a singular behavior, with jump discontinuities, at nontrivial values of the applied field. The lines represent the solution of the exact self-consistent equations (both the physical and unphysical branches are shown) and the symbols represent the simulation results ($N_{sites}=10^5$, $N_{realizations}=2.10^5$). A small finite-size effect is still visible in the simulation data in the jump regions.}
\end{figure}
\end{center}
>From the solution of the coupled set of equations, it is straightforward to compute the average probabilities $\bar{p}_{\tilde{S},S}(x)$. We find that the behavior drastically changes at a critical value of the random-field strength $\sigma$. At large disorder, the set of equations has a unique solution and the average conditional probabilities $\bar{p}_{\tilde{S},S}$ are continuous functions of the applied field, as illustrated in Figure 9. At weak disorder, several branches of solutions are present. As explained above, two of them are ``stable''. One corresponds to the physical process in which the magnetization always increases with the applied field $H$: the average conditional probabilities show a jump discontinuity as one increases $H$ as illustrated in Fig. 10. The ``unphysical'' solution lies in a small range of fields.  This solution is expected to be connected to the lower  branch of the physical solution, but this intermediate portion is unreachable by the iterati!
 on algorithm (hence, is not shown in Fig. 10). We recall that this only concerns one side of the hysteresis loop, the other one being obtained by symmetry.

This behavior is the signature of a disorder-induced out-of-equilibrium transition. Indeed, the magnetization which is given by Eqs. (7) and (8) shows a discontinuous jump as the applied field is increased only when the conditional probabilities are themselves discontinuous (for a value of $x$ different form $-J,J$ or $3J$). The critical value of the disorder separating the two regimes, with and without a discontinuity, is equal to $\sigma_{c}\approx1.006$, as illustrated for $\bar{p}_{+2,+}(H)$ in Figure 11.
\begin{center}
\begin{figure}[th]
\includegraphics[width=7cm,clip]{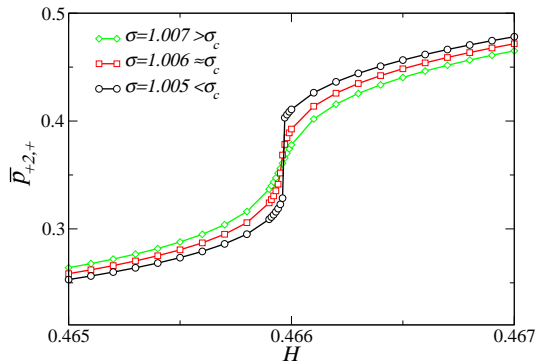}
\caption{\label{fig11} (Color on line) Variation of  the average probability $\bar{p}_{+2,+}(H)$ with disorder strength $\sigma$ for $z=3$ (physical branch). The critical disorder strength is found to be $\sigma_c \simeq 1.006$. The corresponding critical value of the applied field is $H_c \simeq 0.466$. }
\end{figure}
\end{center}

\section{Discussion of the results and comparison with  simulation}
\label{simulation}
Before discussing the above results in more detail, we present the simulation study that we have carried out in parallel. Following the now standard procedure\cite{DSS1997, MP2001}, we have mimicked a Bethe lattice by random graphs of fixed connectivity. These graphs are easily handled in simulations because, even with a finite size, they have no boundary effects, contrary to the Bethe-lattice/Cayley-tree approach. However, the price to pay is that the random graphs have some loops. The (expected) equivalence between the two approaches stems from the fact that the typical loops found in random graphs have a length that scales like the logarithm of the total number of sites.

The simulations are performed on systems of size $N_{sites}=10^4-10^5$. For a Bethe lattice, such sizes are
large enough to avoid significant finite-size effects and the results may be considered to be very close to the thermodynamic limit\cite{IOV2005}. 
In order to compute the magnetization as a function of the external field $H$,  the latter is changed by a fixed amount  $\Delta H=\pm 0.01$. For a given realization $\{h_i\}$ of the random 
fields and a given value of $H$, we relax the system until all pairs are $2$-spin-flip stable, and we then compute the magnetization  $m=\sum_{i=1}^{N} S_i / N $. Similarly, to compute the conditional probabilities, we relax the system under the appropriate constraints. Finally, the results are averaged over $N_{realizations}=10^2-10^5$  different realizations of random graphs and random fields.

A comparison of the exact and simulation results is shown in Figures 6 (average conditional probabilities) and 7 (magnetization) for the one-dimensional chain ($z=2$) as well as in Figures 9 and 10 (average conditional probabilities) for the case $z=3$. The agreement between simulation and theory is excellent. In the case $z=3$ we actually exploit the agreement found for the average conditional probabilities to skip the analytical calculation of the magnetization (a feasible but very tedious task), and discuss instead the simulation results.
\begin{center}
\begin{figure}[th]
\includegraphics[width=7cm,clip]{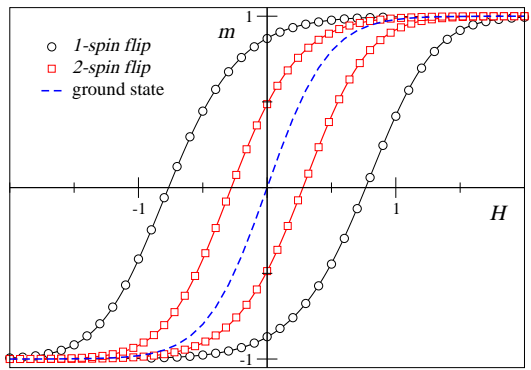}
\caption{\label{fig12} (Color on line) Hysteresis loop, $m(H)$, for the one-dimensional chain  with the $1$- and  $2$-spin-flip dynamics. The lines correspond to the exact results and the symbols to the simulation results ($N_{sites}=10^4$, $N_{realizations}=10^2$). The dashed line is the equilibrium magnetization curve. The disorder strength is $\sigma=1.0$. }
\end{figure}
\end{center}

The influence of the dynamics on the magnetization curve $m(H)$ is illustrated in Figures 12 and 13. In Figure 12, we display the exact results obtained with the 1-spin-flip and 2-spin-flip dynamics for the one-dimensional chain. The width of the hysteresis loop decreases as one goes from 1-spin flips to  2-spin flips and it vanishes for the equilibrium curve (this latter is obtained by solving the exact recursion equations given in Ref.\cite{B1984}). A similar effect is observed for  $z=3$ : see Figure 13 where we have also added the curve obtained with a 3-spin-flip dynamics (whose precise definition will be given below). This effect is quite understable from a physical point of view. Introducing more cooperativity in the dynamics allows the system to avoid being trapped in high-energy minima that are only stable with respect to single spin-flips. As a result, a better ``equilibration'' takes place and hysteresis is reduced.
\begin{center}
\begin{figure}[th]
\includegraphics[width=7cm,clip]{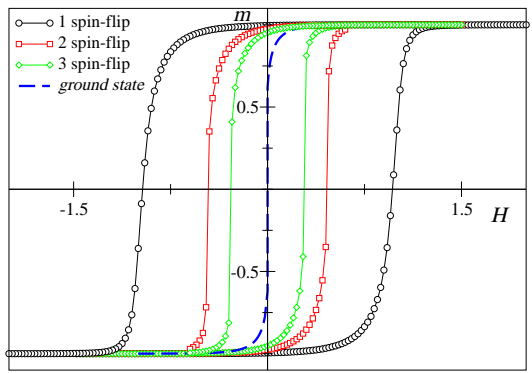}
\caption{\label{fig13} (Color on line) Hysteresis loop, $m(H)$, for the $z=3$ Bethe lattice with the $1$- , $2$-, and $3$-spin-flip dynamics, as well as equilibrium magnetization curve (dashed line):  simulation results for $\sigma=1.0$ ($N_{sites}=10^5$, $N_{realizations}=10^2$). Note the jumps for the $2$-, $3$-spin-flip and equilibrium cases and the continuous behavior in the $1$-spin-flip case.}
\end{figure}
\end{center}

An interesting outcome of the present study is that the peculiar behavior of the hysteresis on a Bethe lattice with connectivity $z=3$ found with the 1-spin-flip dynamics disappears as one considers more cooperative dynamical rules. As discussed in the previous section, an out-of-equilibrium transition is indeed found with the 2-spin-flip dynamics. One can see from Figure 13 that not only does the hysteresis loop shrink as the dynamics is more cooperative, but it also becomes steeper; for the value of the disorder strength considered, it shows a discontinuity for the 2- and 3-spin-flip dynamics (and for the equilibrium curve), whereas it stays continuous for the 1-spin-flip dynamics.

The critical values of the disorder separating continuous from discontinuous behavior satisfy the following inequalities,

\begin{equation}
\sigma_{c}^{1SF}=0<\sigma_{c}^{2SF}\simeq1.006<\sigma_{c}^{EQ}\simeq1.05,
\end{equation}
where the superscripts $1SF$, $2SF$, and $EQ$ refer to the 1-, 2-spin-flip dynamics and to the  equilibrium situation ($\sigma_{c}^{EQ}$ is estimated numerically by solving the recursion equations of Ref.\cite{B1984} for different values of the disorder strength). A similar set of inequalities was found in the computer simulation study of the RFIM on a 3-dimensional cubic lattice, with however a nonzero value of $\sigma_{c}^{1SF}$\cite{VRT2005}. This suggests that increasing the cooperativity in the dynamics of the driven RFIM at $T=0$, on top of reducing hysteresis effects, favors large-scale collective behavior of the spins. Accordingly, the critical value of the disorder above which long-range correlations disappear increases with cooperativity. The absence of transition observed with the 1-spin-flip dynamics for $z=3$ does not appear of physical significance.

It is interesting at this point to discuss in more detail the way to increase the cooperativity in the spin dynamics. The idea is to introduce a $k$-spin-flip dynamics\cite{PV2004,VRT2005}, in which, as already seen for the 1- and 2-spin-flip dynamics, $k$-stable states are visited. The main idea is that the system jumps from one $k$-stable state to the nearest one whenever the former becomes unstable to the simultaneous flip of any subset of $k$ or less spins. For the one-dimensional chain, implementing the rule is rather straighforward: one has to check the stability of continous strings of $k'$ neighboring spins with $k'\leq k$. An exact description of the hysteresis behavior can again be derived. The relevant conditional probabilities now involve the configuration of a string formed by a spin in a given local field and the $k-1$ spins $S_{1},...,S_{z-1}$ at the lower levels, given that the parent spin is down and given the states of the $k-1$ spins that are the descendan!
 ts of spin $S_{z-1}$. The recursion relation that generalizes Eq. (9) requires transition rates for the change of configuration of a string of $k$ spins, knowing the states of the parents and of the descendants. Writting down the equations is formally easy, the difficulty coming with the derivation of the explicit expressions for the transition rates. We therefore do not pursue here in this direction. Nonetheless, we point out that in the limit where $k\rightarrow \infty$ ($k$ goes as the size of the system, here in one dimension), one should likely recover the recursion expressions for the ground state of the RFIM\cite{B1984}. Indeed, the $k$-stable states explored by the system converge now to the ground state\cite{BM2000,MP2003} (expected to be unique for a Gaussian distribution of the random fields).

In higher dimension or for Bethe lattices of higher connectivity, one has to worry about the geometry of the subset of $k'$ spins. In the case of the 3-spin-flip dynamics on a Bethe lattice of connectivity $z=3$, a simple implementation is to restrict the ``geometry'' of the triplets of spins considered (all single spins and all pairs of spins being nonetheless checked so that the 3-stable states so obtained are indeed 1- and 2-stable also): we have considered triplets formed by a spin and its two direct descendants, without checking triplets formed by a spin, a ``child'' in a given branch and a ``grand-child'' at one more level away (see Figure 2). Our simulations have been performed according to this rule.

\section{Conclusion}
\label{conclusion}
In this work, we have obtained the exact solution for the hysteresis behavior of the $T=0$ RFIM on a Bethe lattice when driven by an adiabatically changing applied field and evolved under a 2-spin-flip dynamics. This solution generalizes the results already derived for the 1-spin-flip dynamics. Although we have focused on the main hysteresis loop, it is most likely that further exact results are obtainable for other quantities such as minor hysteresis loops and the avalanche size distribution, albeit with considerable technical difficulty.

Our exact results confirm the robustness of the whole picture concerning the nonequilibrium response of the RFIM under a change of spin dynamics\cite{SDP2004}, robustness that was first shown by a simulation of the RFIM on a cubic lattice\cite{VRT2005}. We find that for lattices with connectivity $z\geq3$ the driven dynamics is characterized by an out-of-equilibrium critical point separating a strong-disorder regime with a smooth hysteresis curve and microscopic avalanches from a weak-disorder regime with a discontinuity in the hysteresis curve associated to a macroscopic avalanche. The width of the main hysteresis loop and the critical value of the disorder depend on the degree of cooperativity of the dynamics, but the overall scenario is unchanged. (As one expects the critical behavior on Bethe lattices to be mean-field-like, the critical point in the 2-spin-flip dynamics trivially falls in the same universality class as that of the 1-spin-flip dynamics and that of the equ!
 ilibrium transition.)

The robustness of the main characteristics associated with hysteresis behavior with respect to the spin dynamics suggests a nontrivial organization of the energy landscape. It was shown for the 1-dimensional chain and conjectured for the general case that the main hysteresis loop of the RFIM under 1-spin-flip dynamics is the outer envelope of all the 1-stable states in the magnetization versus applied-field diagram\cite{DRT2005}. A similar calculation could probably be undertaken for the 2-spin-flip dynamics, but in the absence of such a result, we merely speculate that the main hysteresis loop under $k$-spin-flip dynamics is the envelope of all the $k$-stable states. On physical ground, and as already shown here for the 2- and 3-spin-flip dynamics, one expects the width of the loop to shrink as the dynamics becomes more cooperative. One can then picture the organization of the metastable stable states as formed by a succession of envelopes, the $k$th envelope being included!
  in the $(k-1)$th and so on, until hysteresis vanishes and the envelope collapses to a curve that coincides with the equilibrium (ground state) behavior\cite{note1}. As discussed in Ref.\cite{DRT2005}, a jump discontinuity associated with a macroscopic avalanche occurs when there is a reentrant part in the corresponding envelope: a jump in the $k$-spin-flip dynamics would then imply a jump in more cooperative dynamics, but not necessarily in less cooperative ones. The apparently surprising result concerning the Bethe lattice with $z=3$ can be interpreted along theses lines: there is a transition for the 2-spin-flip dynamics, and for the 3-spin-flip dynamics and the equilibrium as well, but not for the 1-spin-flip dynamics. More work would be needed to confirm the  general picture.

\appendix

\section{The one-dimensional chain ($z=2$)}

>From Eqs. (11), one can see that the conditional probabilities are constant in the intervals $]-\infty,0]$ and $[2J,+\infty [$, $p_{{\tilde S}S}(x)=\hat{p}_{{\tilde S}S}$, whereas $p_{++}(x)$ and $p_{--}(x)$  have an explicit $x$-dependence in the interval $[0,2J]$.
In this interval, the other probabilities are zero and the normalization property implies that 
$p_{++}(x)=1-q(x)$ and $p_{--}(x)=q(x)$, where $q(x)$ is a function to be determined. Introducing the piecewise form of the conditional probabilities in  the right-hand sides of Eqs. (11) leads to the following self-consistent equations:
\begin{align}
&\hat{p}_{++}=p_1-p_0+p_0(\hat{p}_{++}+\hat{p}_{-+})+(p_2-p_1)\hat{p}_{++}\nonumber\\
&\hat{p}_{-+}=(1-p_2)\hat{p}_{+-}+(1-p_1)\hat{p}_{--}\nonumber\\
&\hat{p}_{+-}=p_0(\hat{p}_{++}+\hat{p}_{-+})+\int_0^{2J} \delta x \  (1-q(x))\nonumber\\
&\hat{p}_{--}=(1-p_1)(\hat{p}_{+-}+\hat{p}_{--})+\int_0^{2J} \delta x \  q(x)\nonumber\\
&q(x)=(1-p_1)\hat{p}_{--}+\hat{p}_{+-}\int_{\-\infty}^{-x} \delta x_1 +\int_0^{2J-x} \delta x_1  q(x_1) \ ,
\end{align}
where the $p_n$'s are defined below Eq. (14). The normalization property in the intervals $]-\infty,0]$ and $[2J,+\infty[$ takes the form
\begin{align}
&\hat{p}_{++}+\hat{p}_{-+}=1\nonumber\\
&\hat{p}_{+-}+\hat{p}_{--}=1,
\end{align}
which leads to further simplification of Eqs. (A1) that are reduced to $3$ coupled linear equations, including one inhomogeneous integral equation for $q(x)$.

By using the definition of the average probabilities, Eq. (12), it is easy to rewrite the above equations
in terms of those average probabilities and of $q(x)$, and finally to derive Eqs. (13,14).

Finding the expression for the magnetization, hence for the average probability that the central spin is up, requires the knowledge of the transition rates $ \tilde{W}_{\{ \tilde{S}_{1},\tilde{ S}_{2}|( S'_{1},S'_{2},-1) \rightarrow (S_{1},S_{2},+1)\}}(x_1,x_2,x)$) schematically pictured in Figure 4 (with only two branches merging at the central spin, since $z=2$). The explicit expressions for the  $ \tilde{W}$'s are lengthy and not worth reproducing here, but can be derived in a straightforward manner by studying the stability of the triplet formed by the central spin and its neightbors under the $2$-spin-flip dynamical rule. Inserting the expressions into Eq. (8) (with $z=2$) and using Eqs. (12-14) lead to the following result:
\begin{align}
\bar{P}(+1,|H)&=p_2(\bar{p}_{++}+\bar{p}_{-+})^2+2p_1(\bar{p}_{++}+\bar{p}_{-+})(\bar{p}_{+-}+\bar{p}_{--})+p_0(\bar{p}_{+-}+\bar{p}_{--})^2\nonumber\\
&+2\frac{\bar{p}_{+-}}{1-p_1}[F_{11}(\bar{p}_{++}+\bar{p}_{-+})+F_{10}(\bar{p}_{+-}+\bar{p}_{--})]-(\frac{\bar{p}_{+-}}{1-p_1})^2\tilde{F}_{11}\nonumber\\
&+2(\bar{p}_{++}+\bar{p}_{-+})\int_0^{2J} \delta x \  q(x)\int_{-x}^{0} \delta x_1+2(\bar{p}_{+-}+\bar{p}_{--})\int_0^{2J} \delta x \  q(x)\int_{2J-x}^{2J} \delta x_1\nonumber\\
&-2\frac{\bar{p}_{+-}}{1-p_1}\int_0^{2J} \delta x \  q(x)\int_{2J-x}^{2J} \delta x_1\int_{-x_1}^{0} \delta x_2 -2\int_0^{2J} \delta x \  q(x)\int_{0}^{x} \delta x_1\int_{2J-x_1}^{2J} \delta x_2
\end{align}
where on top of the $p_n(H)$'s associated with $1$-spin flips, we have introduced additional integrals over the (gaussian) distribution of the random field: the functions
\begin{align}
 F_{00}(H)&=\int_0^{2J} \delta x \int_{2J-x}^{2J} \delta x_1\nonumber\\
F_{10}(H)&=F_{01}(H)=\int_0^{2J} \delta x \int_{-x}^{0} \delta x_1\nonumber\\
F_{11}(H)&=\int_{-2J}^{0} \delta x \int_{-(2J+x)}^{0} \delta x_1
\end{align}
represent the probabilities that a pair of neighbors flips up in an irreducibly cooperative way, given that the neighbors of the pair are both down, one up and one down, and both up, respectively, whereas
\begin{align}
\tilde{F}_{11}(H)=\int_0^{2J} \delta x (\int_{-x}^{0} \delta x_1)^2
\end{align}
is related to the  probability to flip a spin up by an irreducibly cooperative flip of any of the two pairs it belongs to, knowing that the outer neighbors of the pairs are up. Again, the dependence of all these quantities on the applied field $H$ comes through the measure $\delta x$ (see below Eq. (6)).

Notice that the first line of Eq. (A3) is exactly the result obtained with single-spin flips only\cite{S1996}. One can check that by setting $q(x)=(1-p_1)/(1-p_1+p_0)$ over the whole interval $[0,2J]$ and using Eqs. (13), the other contributions cancel altogether.

The magnetization is finally obtained through Eq. (7).

\section{ Self-Consistent equations for the case $z=3$}

The transition rates $W_{\left\lbrace \tilde{\bf S}|(S'_1,S'_2,-1) \rightarrow (S_1,S_2,S)|-1\right\rbrace }(x_1,x_2,x)$ corresponding to the case $z=3$ in Figure 3 are obtained by combining two stability diagrams similar to those of Figure 1 in a 3-dimensional graph with axes $x_1,x_2$ and $x$. Deriving the expressions is a tedious task, but with no conceptual difficulty. The expressions lead to the piecewise form of the conditional probabilities given in Eqs. (16) with $4$ different intervals of $x$ to be considered: $]-\infty,-J], [-J,J], [J,3J]$ and $[3J,+\infty[$. The  probabilities are independent of $x$ in the intervals $]-\infty,-J]$ and $[3J,+\infty[$ with values ${\hat p}_{{\tilde S}S}$ (see Eqs. (16)) given by the self-consistent equations
\begin{align}
{\hat p}_{+2,+}&=\left[ \sum_{{\tilde S}}\int_{-J({\tilde S}+1)}^{-J({\tilde S}-1)}\delta x \   p_{{\tilde S},-}(x) +\sum_{{\tilde S}}\int_{-J({\tilde S}-1)}^{\infty}\delta x \   p_{{\tilde S},+}(x) \right]^2\nonumber\\
{\hat p}_{0,+}&=2\left[ \sum_{{\tilde S}}\int_{-\infty}^{-J({\tilde S}+1)}\delta x \  p_{{\tilde S},-}(x)\right] 
\left[\sum_{{\tilde S}}\int_{-J({\tilde S}+1)}^{-J({\tilde S}-1)}\delta x \  p_{{\tilde S},-}(x)+\sum_{{\tilde S}}\int_{-J({\tilde S}-1)}^{\infty} \delta x \  p_{{\tilde S},+}(x) \right]\nonumber\\
{\hat p}_{-2,+}&=\left[ \sum_{{\tilde S}}\int_{-\infty}^{-J({\tilde S}+1)}\delta x \  p_{{\tilde S},-}(x)\right] ^2\nonumber\\
{\hat p}_{+2,-}&=\left[ \sum_{{\tilde S}}\int_{-J({\tilde S}-1)}^{\infty}\delta x \  p_{{\tilde S},+}(x)\right] ^2\nonumber\\
{\hat p}_{0,-}&=2\left[ \sum_{{\tilde S}}\int_{-\infty}^{-J({\tilde S}-1)}\delta x \  p_{{\tilde S},-}(x)\right] \left[ \sum_{{\tilde S}}\int_{-J({\tilde S}-1)}^{\infty}\delta x \  p_{{\tilde S},+}(x)\right]\nonumber\\
{\hat p}_{-2,-}&=\left[ \sum_{{\tilde S}}\int_{-\infty}^{-J({\tilde S}-1)}\delta x \  p_{{\tilde S},-}(x)\right] ^2 \ .
\end{align}

Note that sums and integrals do not commute in the above equations. In the intermediate intervals $[-J,J]$ and $[J,3J]$, the probabilites are functions of $x$ (except $p_{-2,-}$ in $[-J,J]$). The corresponding functions, introduced in Eqs. (16), satisfy the following equations:
\begin{align}
 q_2(x)&=\left[\sum_{{\tilde S}}\int_{-J({\tilde S}-2)-x}^{-J({\tilde S}-1)}\delta x_1 p_{{\tilde S},-}(x_1) \right]\left[2\sum_{{\tilde S}}\int_{-J({\tilde S}+1)}^{-J({\tilde S}-1)}\delta x_1 p_{{\tilde S},-}(x_1)
-\sum_{{\tilde S}}\int_{-J({\tilde S}-2)-x}^{-J({\tilde S}-1)}\delta x_1 p_{{\tilde S},-}(x_1)\right]&\nonumber\\
&+\left[\sum_{{\tilde S}}\int_{-J({\tilde S}-1)}^{\infty}\delta x_1 p_{{\tilde S},+}(x_1) \right]\left[2\sum_{{\tilde S}}\int_{-J({\tilde S}+1)}^{-J({\tilde S}-1)}\delta x_1 p_{{\tilde S},-}(x_1)
+\sum_{{\tilde S}}\int_{-J({\tilde S}-1)}^{\infty}\delta x_1 p_{{\tilde S},+}(x_1)\right]\nonumber\\
q_1(x)&=2\left[ \sum_{{\tilde S}}\int_{-\infty}^{-J({\tilde S}+1)}\delta x_1 p_{{\tilde S},-}(x_1)\right] 
\left[\sum_{{\tilde S}}\int_{-J({\tilde S}-2)-x}^{-J({\tilde S}-1)}\delta x_1 p_{{\tilde S},-}(x_1)+
\sum_{{\tilde S}}\int_{-J({\tilde S}-1)}^{\infty} \delta x_1 p_{{\tilde S},+}(x_1) \right]\nonumber\\
q_0(x)&=\left[ \sum_{{\tilde S}}\int_{-\infty}^{-J({\tilde S}-2)-x}\delta x_1 p_{{\tilde S},-}(x_1)\right] ^2\nonumber\\
r_2(x)&=2\left[ \sum_{{\tilde S}}\int_{-J{\tilde S}-x}^{-J({\tilde S}-1)}\delta x_1 p_{{\tilde S},-}(x_1)\right] \left[\sum_{{\tilde S}}\int_{-J({\tilde S}-1)}^{\infty}\delta x_1 p_{{\tilde S},+}(x_1)\right]+\left[\sum_{{\tilde S}}\int_{-J({\tilde S}-1)}^{\infty} \delta x_1 p_{{\tilde S},+}(x_1)\right]^2\nonumber\\
r_1(x)&=2\left[ \sum_{{\tilde S}}\int_{-\infty}^{-J{\tilde S}-x}\delta x_1 p_{{\tilde S},-}(x_1)\right] \left[ \sum_{{\tilde S}}\int_{-J({\tilde S}-1)}^{\infty}\delta x_1 p_{{\tilde S},+}(x_1)\right] \ .
\end{align}

Note that the $x$-dependence appears through the bounds of the integrals. In Eqs. (B1) and (B2) the conditional probabilities present in the right-hand sides should be replaced by their expressions given by Eqs. (16), which makes the whole set of equations self-consistent. Due to the normalization conditions there are only  $7$ independent coupled equations. All the equations are nonlinear, and those for the $3$ independent functions are inhomogeneous integrals equations as well. The procedure to solve the set of equations is discussed in the main text.

Finally, we point out that, as in the $z=2$ case, the result of the 1-spin-flip dynamics can be recovered by assuming that the functions in the intermediate range $[-J,J]$ and $[J,3J]$ are constant. The proof is a little more complicated than for $z=2$ and is not worth reproducing here.

\section{Acknowledgements}

We acknowledge fruitful discussions with P. Shukla and E. Vives.
This work has   received   financial   support from projects
MAT2004-01291 (CICyT, Spain) and SGR-2001-00066 (Generalitat de
Catalunya).  X.I. acknowledges a grant from DGI-MEC (Spain)
and the hospitality of LPTMC. LPTMC is UMR 7600 at the CNRS.


\begin{thebibliography}{10}
\bibitem{note1} One should recall at this point that a $k$-stable state is also stable against the flips of smaller numbers of spins.
\end{thebibliography}
\end{document}